\documentclass{easychair}

\pdfoutput=1

\newcommand{\xf}[1]{Figure~\ref{#1}}

\newcommand{\xl}[1]{Listing~\ref{#1}}

\newcommand{\gipc}{{GIPC\index{GIPC}\index{Frameworks!GIPC}}}

\newcommand{\gee}{{GEE\index{GEE}\index{Frameworks!GEE}}}

\newcommand{\gipsy}{{GIPSY\index{GIPSY}}}
\newcommand{\ripe}{{RIPE\index{RIPE}\index{Frameworks!RIPE}}}

\newcommand{\gipl}{{GIPL\index{GIPL}}}

\newcommand{\lucid}{{Lucid\index{Lucid}}}

\newcommand{\jlucid}{{JLucid\index{JLucid}}}
\newcommand{\olucid}{{Objective Lucid\index{Tensor Lucid}}}

\newcommand{\flucid}{{Forensic Lucid\index{Forensic Lucid}}}

\newcommand{\lucx}{{Lucx\index{Lucx}}}

\newcommand{\jooip}{{JOOIP\index{JOOIP}}}

\newcommand{\java}{{Java\index{Java}}}

\newcommand{\trans}{$\psi$}

\newcommand{\invtrans}{$\Psi^{-1}$}

\newcommand{\api}[1]{\texttt{#1}\index{API!#1}}

\newcommand{\lucidL}[1]{{$\mathit{Lucid}$}($L$) }

		{}

\def\myvert{\raise 2.27pt \hbox{\vrule depth 0pt height 8pt width 0.2mm}}
\def\myarrow{\hspace*{0.43mm}%
             \raise 2.29pt\hbox{\vrule depth 0pt height 8pt width 0.16mm}%
             \hspace*{-0.32mm}%
             $\longrightarrow$
             \ %
             }

\lstdefinestyle{codeStyle}
{
        language=Java,
        frame=single,  
        basicstyle=\footnotesize,
        captionpos=b,
        showstringspaces=false,
        showspaces=false,
        extendedchars=true,
        linewidth=1\linewidth,
        breaklines=true,
        float=phtb  
}

\begin{document}

%
%

\title{Towards Automated Deduction in Blackmail Case Analysis with Forensic Lucid}
\titlerunning{Automated Deduction in Blackmail Case Analysis with Forensic Lucid}

\author{Serguei A. Mokhov \hspace{1cm} Joey Paquet \hspace{1cm} Mourad Debbabi\\
Faculty of Engineering and Computer Science\\
Concordia University, Montr\'eal, Qu\'ebec, Canada,\\
\url{{mokhov,paquet,debbabi}@encs.concordia.ca}%
}

\authorrunning{Mokhov et al.}

\maketitle

\begin{abstract}
This work-in-progress focuses on the refinement of application of the intensional
logic to cyberforensic analysis and its benefits are compared with the finite-state automata approach.
This work extends the use of the scientific intensional programming paradigm onto modeling and
implementation of a cyberforensics investigation process with the backtrace of event reconstruction,
modeling the evidence as multidimensional hierarchical contexts, and proving or disproving the
claims with it in the intensional manner of evaluation.
This is a practical, context-aware improvement over the finite state automata (FSA) approach we have seen in
the related works.
As a base implementation language model we use in this approach is a new
dialect of the Lucid programming language, that we call {\flucid} and in this paper
we focus on defining hierarchical contexts based on the intensional logic for the
evaluation of cyberforensic expressions.
\end{abstract}

\section{Introduction}

\paragraph*{Problem Statement.}

The first formal approach for event reconstruction
cyberforensic analysis
appeared in two papers~\cite{printer-case,blackmail-case}
by Gladyshev et al. that relies on the finite-state automata (FSA)
and their transformation and operation to model evidence, witnesses,
stories told by witnesses, and their possible evaluation. One of the examples the papers present
is the use-case for the proposed technique -- Blackmail Investigation.
We aim at the same case to model and implement it
using the new approach, which promises to be more friendly and usable in the actual investigator's work and
serve as a basis to further development in the area.

\paragraph*{Proposed Solution.}

We intend to show the intensional approach with a {\lucid}-based dialect
to the problem is an asset in the field of cyberforensics as it
is promising to be more practical and usable than the FSA.
Since {\lucid} was originally designed and used to prove correctness of programming
languages~\cite{lucid76,lucid77}, and is based on the temporal logic, functional and data-flow languages
its implementation to backtracking in proving or disproving the evidential statements and claims in the
investigation process as a evaluation of an expression that either evaluates to {\em true} or {\em false} given all
the facts in the formally specified context. We will also attempt to retain the generality of the approach vs. building
a problem-specific FSA in the FSA approach that can suffer a state explosion problem.

From the logic perspective, it was shown one can model computations (the basic unit in the finite state machines
in~\cite{printer-case,blackmail-case}) as logic~\cite{lalement-plaice93}. When armed with contexts as first-class
values and a demand-driven model
adopted in the implementation of the Lucid-family of languages~\cite{gipsy2005,gipsy-simple-context-calculus-08,gipsy-multi-tier-secasa09,kaiyulucx,tongxinmcthesis08} that limits the scope of evaluation in a given set of
dimensions, we come to the intensional logic and the corresponding programming artifact. In the essence, we model our forensic
computation unit in the intensional logic and propose
the ways to implement it in practice within an intensional programming
platform~\cite{gipsy,gipsy2005,mokhovmcthesis05,gipsy-multi-tier-secasa09}. We see a lot of potential for this work to be successful and beneficial for
cyberforensics as well as intensional programming communities.

Based on the parameters and terms defined in the papers~\cite{printer-case,blackmail-case}, we have various pieces of evidence and witnesses telling their own
stories of the incident. The goal is to put them together to make the description of the incident as precise as possible. To
show that a certain claim may be true, an investigator has to show that there are some explanations of evidence that agrees
with the claim. To disprove the claim, the investigator has to show there is no explanation of evidence that agree with the
claim~\cite{printer-case}.
The authors of the FSA approach did a proof-of-concept implementation of the algorithms
in CMU Common LISP~\cite{printer-case} that we target to improve by re-writing in a Lucid dialect,
that we call {\flucid}. In this work we focus on the specification of hierarchical context expressions
and the operators on them when modeling the examples. LIPS, unlike {\lucid}, entirely
lacks contexts build into its logic, syntax, and semantics, thereby making the
implementation of the cases more clumsy and inefficient (i.e. highly sequential).
Our system~\cite{gipsy} (not discussed here) offers distributed demand-driven evaluation of Lucid
programs in a more efficient way and is more general than LISP's compiler and run-time environment.

\paragraph*{Lucid Overview.}

{\lucid}~\cite{lucid76,lucid77,lucid85,lucid95,nonprocedural-iterative-lucid-77} is a dataflow intensional and functional programming language. In fact, it is a family of languages that are built upon intensional logic (which in turn can be understood as a multidimensional generalization of temporal logic) involving context and demand-driven parallel computation model. A program written in some {\lucid} dialect is an expression that may have subexpressions that need to be evaluated at certain {\it context}. Given the set of dimensions $D=\{dim_i\}$ in which an expression varies, and a corresponding set of indexes or {\it tags} defined as placeholders over each dimension, the context is represented as a set of $<\!\!dim_i:tag_i\!\!>$ mappings and each variable in {\lucid}, called often a {\em stream}, is evaluated in that defined context that may also evolve using context operators~\cite{gipsy-simple-context-calculus-08,kaiyulucx,tongxinmcthesis08,wanphd06}. The generic version of Lucid, the General Intensional Programming Language ({\gipl})~\cite{paquetThesis}, defines two basic operators @ and \# to navigate (switch and query) in the contexts $\mathcal{P}$.
The {\gipl} is the first\footnote{The second being {\lucx}~\cite{kaiyulucx,tongxinmcthesis08,wanphd06}} generic programming language of all intensional languages, defined by the means
of only two intensional operators \texttt{@} and \texttt{\#}. It has been proven that other intensional programming languages of the Lucid family can be translated into the {\gipl}~\cite{paquetThesis}.

%
%

\subsection{General Intensional Programming System (GIPSY).}

The {\gipsy}~\cite{gipsy-arch-2000,gipsy,bolu04,wuf04,mokhovmcthesis05,dmf-plc05,wu05,gipsy2005,gipsy-multi-tier-secasa09} is a platform implemented primarily in {\java} to investigate properties of the Lucid family of languages and beyond. It executes {\lucid} programs following a demand-driven distributed generator-worker architecture, and is designed as a modular collection of frameworks where components related to the development ({\ripe}\footnote{Run-time Integrated Programming Environment, implemented in \api{gipsy.RIPE}}), compilation ({\gipc}\footnote{General Intensional Programming Compiler, implemented in \api{gipsy.GIPC}}), and execution ({\gee}\footnote{General Eduction Engine, implemented in \api{gipsy.GEE}}) of Lucid programs are separated allowing easy extension, addition, and replacement of the components.
This is a proposed testing and investigation platform for our {\em Forensic Lucid} language.

\section{{\flucid} Overview}
\label{sect:flucid}
\label{sect:forensic-lucid}

This section summarizes
concepts and considerations in the design of the {\flucid} language, which
is being studied through another
use-case than related works~\cite{flucid-isabelle-techrep-tphols08,flucid-imf08}.
The end goal is to define our {\flucid} language where its constructs concisely
express cyberforensic evidence as context of evaluations, which can be initial state of the case
towards what we have actually observed as a final state in the FSM.
The implementing system~\cite{gipsy2005,gipsy,mokhovmcthesis05,gipsy-multi-tier-secasa09} backtraces intermediate results
to provide the corresponding event reconstruction path
if it exists (which we do not discuss in this work).
The result of the expression in its basic form is either {\em true} or {\em false},
i.e. ``guilty'' or ``not guilty'' given the evidential evaluation context per explanation with the backtrace. There
can be multiple backtraces, that correspond to the explanation of the evidence (or lack thereof).

\paragraph*{Properties.}
\index{{\flucid}!Properties}

We define {\flucid} to model the evidential statements and other expressions representing the evidence and observations as context. An execution trace of a running {\flucid} program is designed to expose the possibility of the proposed claim with the events that lead to the conclusion. {\flucid} aggregates the features of multiple
Lucid dialects mentioned earlier needed for these tasks along with its own extensions.
The addition of the context calculus from {\lucx} (stands for ``Lucid enriched with context'' that promotes contexts as first-class values) for operators on simple contexts and context sets (\api{union}, \api{intersection}, etc.) are used to manipulate complex hierarchical context spaces in {\flucid}. Additionally, {\flucid} inherits many of the properties of {\lucx}, {\olucid}, {\jooip} (Java-embedded Object-Oriented Intensional Programming language), and their comprising dialects, where the former is for the context calculus, and the latter for the arrays and structural representation of data for modeling the case data structures such as events, observations, and groupings of the related data, and so on.
(We eliminate the OO-related aspects from this work as well as some others to conserve space
and instead focus on the context hierarchies, syntax, and semantics.)
Hierarchical contexts are also following the example of MARFL~\cite{marfl-context-secasa08} using a dot operator
and by overloading @ and \# to accept different types as their left and right arguments.
One of the basic requirements is that the final target definition of the syntax, and the operational semantics of {\flucid} should be compatible with the basic {\lucx} and {\gipl}. This is necessary for compiler and and the run time system within the implementing system, called General Intensional Programming System (GIPSY)~\cite{gipsy2005,gipsy,gipsy-multi-tier-secasa09}. The translation rules or equivalent are to be provided when implementing the language compiler within {\gipsy}, and such that the run-time environment (General Eduction Engine, or {\gee}) can execute it with minimal changes to {\gee}'s implementation.

\paragraph*{Context.}
\index{Context}
\index{{\flucid}!Context}

We need to provide an ability to encode the stories told by
the evidence and witnesses. This will constitute the context
of evaluation. The return value of the evaluation would be a collection
of backtraces, which contain the ``paths of truth''. If
a given trace contains all truths values, it's an explanation
of a story. If there is no such a path, i.e. the trace, there
is no enough supporting evidence of the entire claim to be
true.

The context for this task for simplicity of the prototype language can be expressed as integers or strings, to which we attribute some meaning or description.
The contexts are finite and can be navigated through in both directions of the index, potentially allowing negative tags in our tag sets of dimensions.
Concurrently, our contexts can be a finite set of symbolic labels and their values that can internally be enumerated. The symbolic approach is naturally more appropriate for humans and we have a machinery to so in {\lucx}'s implementation in {\gipsy}~\cite{gipsy-context-calculus-07,tongxinmcthesis08}.
We define streams of observations as our context, that can be a simple context or a context set. In fact, in {\flucid} we are defining higher-level dimensions and lower-level dimensions. The highest-level one is the {\em evidential statement}, which is a finite unordered set of observation sequences.
The {\em observation sequence} is a finite {\em ordered} set of observations.
The {\em observation} is an ``eyewitness'' of a particular property along with the duration of the observation.
As in the FSA~\cite{blackmail-case,printer-case}, the observations are tuples of $(P,min,opt)$ in their generic form.
The observations in this form, specifically, the property $P$ can be exploded further into {\lucx}'s context set and further into an atomic
simple context~\cite{wanphd06,gipsy-simple-context-calculus-08}.
Context switching between different observations is done naturally with the {\lucid} @ context switching operator.
Consider some conceptual expression of a storyboard in \xl{list:story-board-expression} where anything in \verb+[ ... ]+ represents a story, i.e. the context of evaluation. \texttt{foo} can be evaluated at multiple contexts (stories), producing a collection of final results (e.g. {\em true} or {\em false}) for each story as well as a collection of traces.

\begin{lstlisting}[
    label={list:story-board-expression},
    caption={Intensional Storyboard Expression},
    style=codeStyle
    ]
foo @
{
  [ final observed event, possible initial observed event ],
  [            ],
  [            ]
}
\end{lstlisting}

While the \verb+[...]+ notation here may be confusing with respect to the notation of \texttt{[dimension:tag]}
in {\lucid} and more specifically in {\lucx}~\cite{wanphd06,tongxinmcthesis08,gipsy-simple-context-calculus-08}, it is in fact a simple
syntactical extension to allow higher-level groups of contexts where this syntactical sugar is later translated
to the baseline context constructs.
The tentative notation of \verb+{[...],...,[...]}+ implies a notion similar to the notion of the ``context set''
in~\cite{wanphd06,gipsy-simple-context-calculus-08,tongxinmcthesis08} except with the syntactical sugar mentioned earlier where we allow
syntactical grouping of properties, observations, observation sequences, and evidential statements as our context sets.
The generic observation sequence~\cite{printer-case}
can be
expanded into the context stream using the $min$ and $opt$ values, where
they will translate into index values. Thus, $obs=(A,3,0)(B,2,0)$ expands
the property labels $A$ and $B$ into a finite stream of five indexed
elements: $AAABB$. Thus, a {\flucid} fragment in \xl{list:duplicate-context-value-tags-code} would return the third
$A$ of the $AAABB$ context stream in the observation portion of $o$. Therefore, possible evaluations to check for
the properties can be as shown in \xf{fig:duration-eval}.
\begin{lstlisting}[
    label={list:duplicate-context-value-tags-code},
    caption={Observation Sequence With Duration},
    style=codeStyle
    ]
// Give me observed property at index 2 in the observation sequence obs
o @.obs 2
where
  // Higher-level dimension in the form of (P,min,opt)
  observation o;
  // Equivalent to writing = { A, A, A, B, B };
  observation sequence obs = (A,3,0)(B,2,0);
  where
    // Properties A and B are arrays of computations
    // or any Expressions
    A = [c1,c2,c3,c4];
    B = E;
    ...
  end;
end;
\end{lstlisting}

The property values of $A$ and $B$ can be anything that context calculus allows.
The \api{observation sequence} is a finite ordered context tag set~\cite{tongxinmcthesis08} that
allows an integral ``duration'' of a given tag property. This may seem like we allow duplicate tag values that are
unsound in the classical {\lucid} semantics; however, we find our way around little further in the text
with the implicit tag index. The semantics of the arrays of computations is not a part of either
{\gipl} or {\lucx}; however, the arrays are provided by {\jlucid} and {\olucid}. We need the notion
of the arrays to evaluate multiple computations at the same context. Having an array of computations
is conceptually equivalent of running an a {\lucid} program under the same context for each array element
in a separate instance of the evaluation engine and then the results of those expressions are gathered
in one ordered storage within the originating program. Arrays in {\flucid} are needed to represent
a set of results, or {\em explanations} of evidential statements, as well as denote some properties
of observations. We will explore the notion of arrays in {\flucid} much greater detail in the near
future work. In the FSA approach computations $c_i$ correspond to the state $q$ and event $i$ that
enable transition. For {\flucid}, we can have $c_i$ as theoretically any Lucid expression $E$.

\begin{figure}[htb!]
\hrule
\small
\begin{verbatim}

Observed property (context): A A A B B
        Sub-dimension index: 0 1 2 3 4

o @.obs 0 = A
o @.obs 1 = A
o @.obs 2 = A
o @.obs 3 = B
o @.obs 4 = B

To get the duration/index position:

o @.obs A = 0 1 2
o @.obs B = 3 4

\end{verbatim}
\normalsize
\hrule
\caption{Handling Duration of an Observed Property in the Context}
\label{fig:duration-eval}
\end{figure}

In \xf{fig:duration-eval} we are illustrating a possibility to query for the sub-dimension indices by raw property
where it persists that produces a finite stream valid indices that can be used in subsequent expressions, or,
alternatively by supplying the index we can get the corresponding raw property at that index. The latter
feature is still under investigation of whether it is safe to expose it to {\flucid} programmers or make
it implicit at all times at the implementation level.
This is needed to remedy the problem of ``duplicate tags'':
as previously mentioned, observations form the context and allow durations. This means multiple
duplicate dimension tags with implied subdimension indexes should be allowed as
the semantics of a traditional Lucid approaches do not allow duplicate
dimension tags. It should be noted however, that the combination of
the tag and its index in the stream is still unique and can be folded
into the traditional Lucid semantics.

\paragraph*{Transition Function.}
\label{sect:tans-func}
\index{Transition Function}
\index{{\flucid}!Transition Function}

A transition function (described at length~\cite{printer-case,blackmail-case} and the derived works)
determines how the context of evaluation changes
during computation. A general issue exists that we have to address
is that the transition function {\trans} is usually problem-specific.
In the FSA approach, the transition function is the labeled graph itself. In the first prototype,
we follow the graph to model our {\flucid} equivalent.
In general,
{\lucid} has already basic operators to navigate and switch from one context to another,
which represent the basic transition functions in themselves (the intensional
operators such as @, \#, \api{iseod}, \api{first}, \api{next}, \api{fby}, \api{wvr}, \api{upon},
and \api{asa} as well as their inverse operators\footnote{Defined further}).
However, a specific problem being modeled requires more specific transition function
than just plain intensional operators. In this case the transition function is
a {\flucid} function where the matching state transition modeled through
a sequence of intensional operators.
A question arises a of how to explicitly model
the transition function {\trans} and its backtrace {\invtrans} in the new language.
A possible approach is to use predefined macros in Lucid syntax~\cite{mokhov-cyberforensics-07}.
In fact, the forensic operators are just pre-defined functions that rely on the traditional
and inverse Lucid operators as well as context switching operators
that achieve something similar to the transitions.
At the implementation level, it is the {\gee} that actually does
the execution of {\trans} within {\gipsy}.
In fact, the intensional operators
of {\lucid} represent the basic building blocks for {\trans} and {\invtrans}.

\paragraph*{Operational Semantics.}
\label{appdx:semantics}
\label{sect:semantics}
\index{Operational Semantics}
\index{Forensic Lucid!Operational Semantics}
\index{Operational Semantics!Forensic Lucid}
\index{Operational Semantics!Indexical Lucid}
\index{Operational Semantics!Lucx}

As previously mentioned, the operational semantics of {\flucid} for the large
part is viewed as a composition of the semantic rules of {\gipl}, {\olucid}, and {\lucx}
along with the new operators and definitions.
The explanation of the rules and the notation
are given in great detail in the cited works and are trimmed in this extended abstract due to shortage of space.
The {\olucid} semantic rules were affected and refined by some of the semantic rules of
{\jooip}~\cite{gipsy-jooip}. We also omit the {\olucid} and {\jooip} semantic rules
due to space limitation and defer them to another publication.
The new rules of the operational semantics of {\flucid} cover the newly defined
operators primarily, including the reverse and logical stream operators as well
as forensic-specific operators. Refining the semantics of context set operators of {\lucx},
such as \api{box} and \api{range} are also a part of this work. We use the same
notation as the referenced languages to maintain consistency in defining our
rules.

\section{Initial Blackmail Case Modeling}
\index{Blackmail Investigation}
\index{Cases!Blackmail}

\begin{figure}[htb!]
	\begin{centering}
	\includegraphics[width=.5\textwidth]{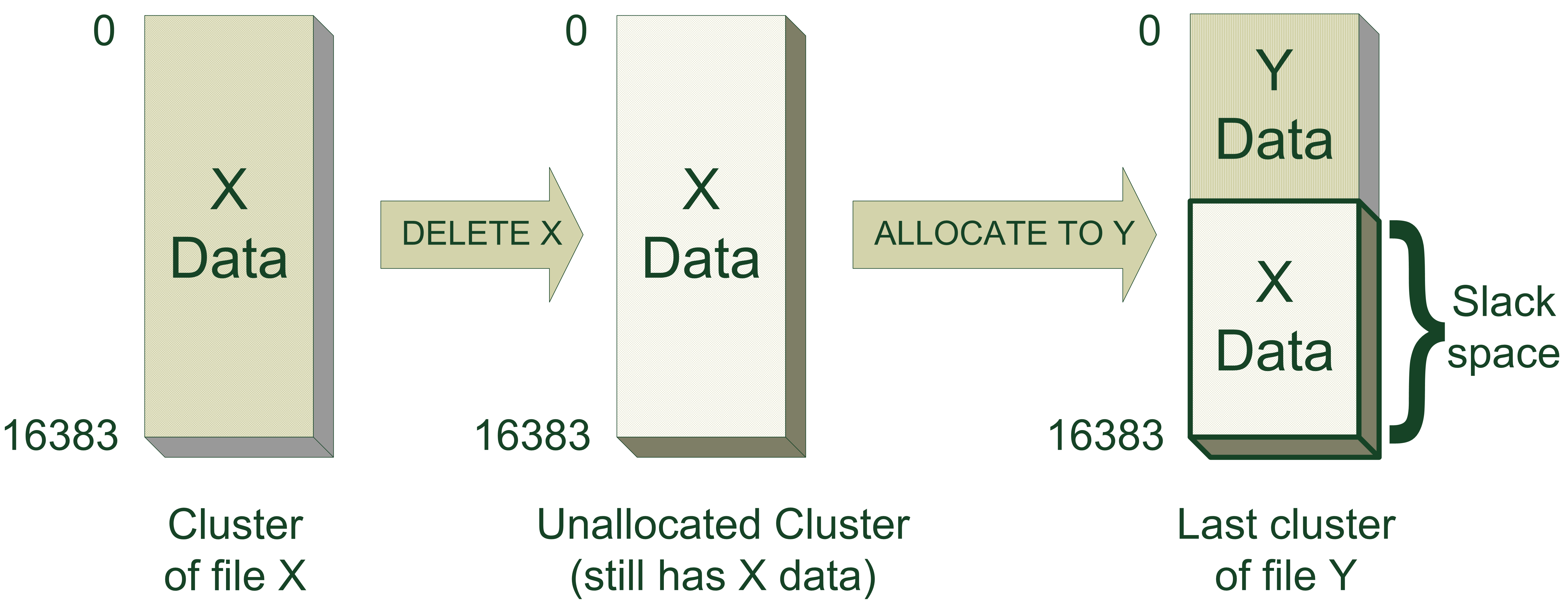}
	\caption{Cluster Data with the Blackmail Fragments}
	\label{fig:blackmail-disk}
	\end{centering}
\end{figure}

The case description in this section is from \cite{blackmail-case}.
A managing director of some company, Mr. C, was blackmailed. He
contacted the police and handed them evidence in the form of a floppy disk
that contained a letter with a number of allegations, threats, and
demands.
The message was known to have come from his friend Mr. A. The police
officers went to interview Mr. A and found that he was on holiday abroad.
They seized the computer of Mr. A and interviewed him as soon as he
returned into the country. Mr. A admitted that he wrote the letter, but denied
making threats and demands. He explained that, while he was on holiday,
Mr. C had access to his computer. Thus, it was possible that Mr. C added the
threats and demands into the letter himself to discredit Mr. A.
One of the blackmail fragments was found in the slack space of another
letter unconnected with the incident. When the police interviewed the
person to whom that letter was addressed, he confirmed that he had
received the letter on the day that Mr. A had gone abroad on holiday. It was
concluded that Mr. A must have added the threats and demands into the
letter before going on holiday, and that Mr. C could not have been involved.
In \xf{fig:blackmail-disk} is the initial view of the incident as a
diagram illustrating cluster data of the blackmail and unconnected letters.

\paragraph*{Modeling the Investigation.}

\begin{figure}[htb!]
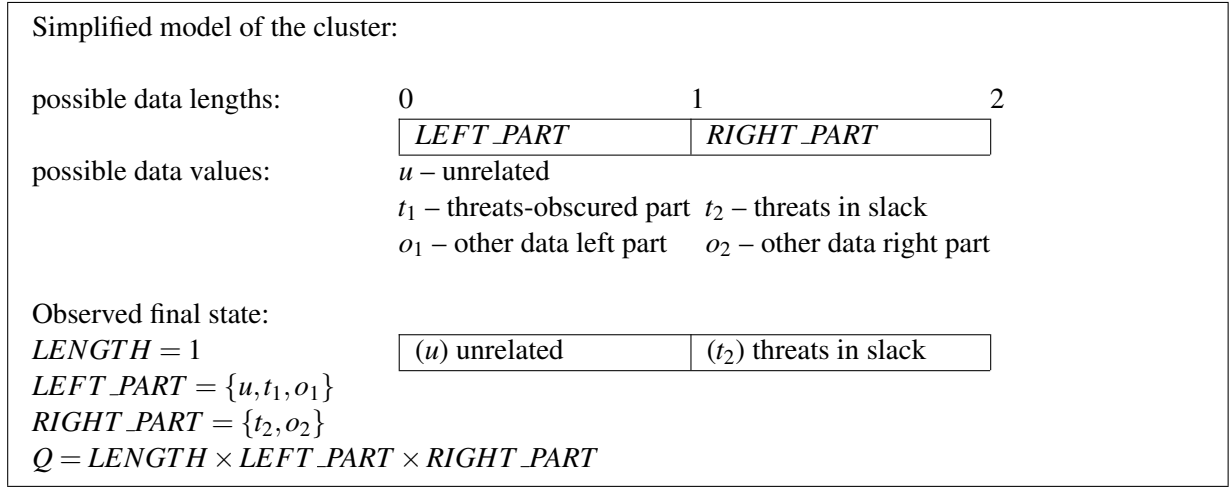

	\begin{centering}
	\fbox{\begin{minipage}{\textwidth}
	\begin{tabular}{l@{}l@{}@{}l@{}@{}l}
	Simplified model of the cluster: &                                    &                                    & \\\\
	possible data lengths:           & 0                                  & 1                                  & 2\\
	\cline{2-3}
	                                 & \vline \; $LEFT\_PART$             & \vline \; $RIGHT\_PART$            & \vline \\
	\cline{2-3}
	possible data values:            & $u$ -- unrelated                   &                                    & \\
	                                 & $t_1$ -- threats-obscured part     & \; $t_2$ -- threats in slack       & \\
	                                 & $o_1$ -- other data left part      & \; $o_2$ -- other data right part  & \\\\
	Observed final state:            &                                    &                                    & \\
	\cline{2-3}
	$LENGTH = 1$                     & \vline \; ($u$) unrelated          & \vline \; ($t_2$) threats in slack & \vline \\
	\cline{2-3}
	$LEFT\_PART = \{u,t_1,o_1\}$     &                                    &                                    & \\
	$RIGHT\_PART = \{t_2,o_2\}$      &                                    &                                    & \\
	\multicolumn{2}{l}{$Q = LENGTH \times LEFT\_PART \times RIGHT\_PART$} &                                    & \\
	\end{tabular}
	\end{minipage}}
	\caption{Simplified Cluster Model}
	\label{fig:blackmail-cluster-simplified}
	\end{centering}
\end{figure}

In the blackmail example, the functionality of the last cluster of a file
was used to determine the sequence of events and, hence, to
disprove Mr. A's alibi.
Thus, the scope of the model can be restricted to the functionality of
the last cluster in the unrelated file.
The last cluster model can store data objects of only three possible
lengths: $LENGTH = \{0,1,2\}$.
Zero length means that the cluster is unallocated.
The length of $1$ means that the cluster contains the object of the size of the
unrelated letter tip.
The length of $2$ means that the cluster contains the object of the size of the
data block with the threats. In \xf{fig:blackmail-cluster-simplified} is,
therefore, the simplified model of the investigation.

\paragraph*{Events.}

The state of the last cluster can be changed by three types of
events:

\begin{enumerate}
\item
	Ordinary writes into the cluster:

	$WRITE=\{(u),(t_1),(o_1),(u,t_2),(u,o_2),(t_1,t_2),(t_1,o_2),(o_1,t_2),(o_1,o_2)\}$

\item
	Direct writes into the file to which the cluster is allocated
	(bypassing the OS):

	$DIRECT\_WRITE=\{d(u,t_2),d(u,o_2),d(o_1),d(t_1,t_2),d(t_1,o_2),d(o_1,t_2),d(o_1,o_2)\}$

\item
	Deletion of the file which sets the length of the file to zero:

	$I = WRITE \bigcup DIRECT\_WRITE \bigcup \{del\}$
\end{enumerate}

\paragraph*{Formalization of the Evidence.}

The final state observed by the investigators is $(1,u,t_2)$.
Let $O_{final}$ denote the observation of this state.
The entire sequence of observations is then $os_{final} = (\$, O_{final})$.
The observation sequence $os_{unrelated}$ says that the unrelated letter
was created at some time in the past, and that it was received by the
person to whom it was addressed is $os_{unrelated} = (\$, O_{unrelated}, \$, (C_T,0,0), \$ )$
where $O_{unrelated}$ denotes the observation that the ``unrelated'' letter tip
$(u)$ is being written into the cluster.
The evidential statement is then: $es_{blackmail} = (os_{final}, os_{unrelated})$.


\paragraph*{Finding an Explanation of Mr. A's Theory.}

Mr. A's theory, encoded using the proposed notation, is  $os_{Mr. A} = (\$,O_{unrelated-clean},\$,O_{blackmail},\$)$,
where $O_{unrelated-clean}$ denotes the observation that the ``unrelated''
letter $(u)$ is being written into the cluster and, at the same time,
the cluster does not contain the blackmail fragment;
$O_{blackmail}$ denotes the observation that the right part of the
model now contains the blackmail fragment $(t_2)$.


\paragraph*{Explanations.}

There are two most logically possible explanations that can be represented by state machine.
See the corresponding state diagram for the blackmail case in \xf{fig:blackmail-states}.

\begin{figure}[htb!]
	\begin{centering}
	\includegraphics[width=0.5\textwidth]{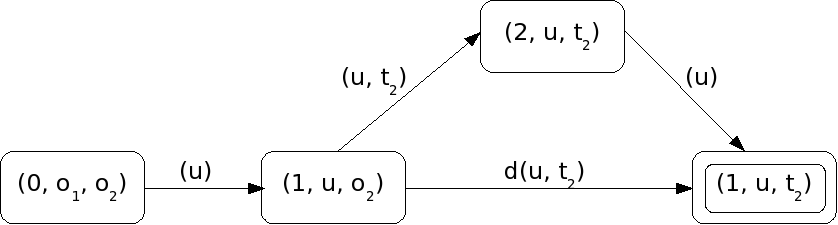}
	\caption{Blackmail Case State Machine}
	\label{fig:blackmail-states}
	\end{centering}
\end{figure}

\begin{enumerate}
\item
The first explanation:

$\ldots \xrightarrow{(u)} (1,u,o_2) \xrightarrow{(u,t_2)} (2,u,t_2) \xrightarrow{(u)} (1,u,t_2)$

\begin{itemize}
\item
	Finding the unrelated letter, which was written by Mr. A earlier;
\item
	Adding threats into the last cluster of that letter by editing it ``in-place''
	with a suitable text editor (such as ViM~\cite{vim});
\item
	Restoring the unrelated letter to its original content by editing it
	``in-place'' again.

\begin{quote}
{\it ``To understand this sequence of events, observe that certain text
editors (e.g. ViM \cite{vim}) can be configured to edit text ``in-place''.
In this mode of operation, the modified file is written back into the
same disk blocks that were allocated to the original file. As a result,
the user can forge the file's slack space by (1) appending the
desired slack space content to the end of the file, (2) saving it,
(3) reverting the file back to the original content, (4) saving it again.''}~\cite{blackmail-case}
\end{quote}

\end{itemize}

\item
The second explanation:

$\ldots \xrightarrow{(u)} (1,u,o_2) \xrightarrow{d(u,t_2)} (1,u,t_2)$

\begin{itemize}
\item
	The threats are added into the slack space of the unrelated
	letter by writing directly into the last cluster using, for example, a
	low-level disk editor.
\end{itemize}
\end{enumerate}

The blackmail case example of the initial implementation steps is in \xl{list:flucid-blackmail}.

\begin{lstlisting}[
    label={list:flucid-blackmail},
    caption={Blackmail Case Modeling in {\flucid}},
    style=codeStyle
    ]
MrA @ es_mra
where
  evidential statement es_mra = {os_mra, os_final, os_unrelated};

  observation sequence os_mra = ($, o_unrelated_clean, $, o_blackmail, $);
  observation sequence os_final = ($, o_final);
  observation sequence os_unrelated = ($, o_unrelated, $, (Ct,0,0), $);

  observation o_final = (1, "u", "t2");
  observation o_unrelated_clean = (1, "u", "o1");

  //...

  invtrans(Q, es_mra, o_final) = backraces
  where
    // list of all possible dimensions
    observation Q = lengths box left_part box right_part;

    // events
    observation lengths = unordered {0, 1, 2};

    // symbolic labels map to human descriptions
    observation left_part = unordered {
        "u"  => "unrelated",
        "t1" => "threats-obscured part",
        "o1" => "other data (left part)"
    };

    observation right_part = unordered {
        "t2" => "threats in slack",
        "o2" => "other data (right part)"
    };

    backtraces = [ A, B, C, D, ];
    where
    ...
    end;
  end;

end;
\end{lstlisting}

\section{Conclusion}

The proposed practical approach in the cyberforensics field
can also be used in a normal investigation process involving
crimes not necessarily associated with information technology.
Combined with an expert system (e.g. implemented in CLIPS~\cite{clips}),
it can also be used in training new staff in investigation techniques.
The focus on hierarchical contexts as first-class values brings more understanding
of the process to the investigators in cybercrime case management tools.

\paragraph*{Future Work.}

\begin{itemize}
\item
	{\flucid} Compiler and run-time environment.

\item
	Prove equivalence to the FSA approach.

\item
  Adapt/re-implement a graphical UI based on the data-flow graph tool~\cite{leitao04} to
  simplify {\flucid} programming for not very tech-savvy
  investigators.
\end{itemize}

\paragraph*{Acknowledgments.}

This research work was funded by the Faculty
of Engineering and Computer Science of Concordia University,
Montreal, Canada.

%
\label{sect:bib}
\bibliographystyle{unsrt}
\bibliography{flucid-blackmail-case}

\end{document}